\documentstyle[preprint,wrapfig,epsf]{ptptex}

\preprintnumber{
        DPNU-95-42\\
        hep-ph/9511339}

\title{%
Dynamical Mass Generation in QED3 with Chern--Simons Term}

\author{%
Pieter {\sc Maris}
\footnote{E-mail: maris@eken.phys.nagoya-u.ac.jp}
}

\inst{%
Department of Physics, Nagoya University, Nagoya 464--01, Japan
}

\recdate{%
\today
}

\abst{%
In (2+1)-dimensional QED with a Chern-Simons term, we study dynamical
breaking of chiral symmetry, using the Dyson--Schwinger equation for
the fermions. There is a region in parameter space were dynamical
chiral symmetry breaking occurs, just as in pure QED3 (without
Chern--Simons term); outside this region, this chiral symmetry
breaking solution does not exists. Our results, both numerically and
analytically, show that the chiral phase transition is a discontinuous
first-order transition.}

\begin{document}

\maketitle

\section{Introduction}

Dynamical symmetry breaking in (2+1)-dimensional QED has attracted a
lot of attention over the last decade\cite{qed3gen,lat}. From a purely
theoretical point of view, it is an interesting and useful theory to
study dynamical mass generation. It also has some applications, not
only in in elementary particle physics, but also in condensed matter
physics, in connection with phenomena occurring in planar surfaces.
For particle physics, the main interest in three-dimensional theories
comes from the fact that they are the high-temperature limit of the
corresponding four-dimensional theory, and thus play a role in studies
of the early universe and the electroweak phase transition.

A natural extension of pure QED3 is to add a Chern--Simons (CS) term
for the gauge field\cite{cs}, which breaks parity explicitly.
Recently, it has been shown that in the presence of such a CS term
there is a {\em first-order} chiral phase
transition\cite{KMpr95,Ko95}. The CS term generates a parity-odd mass
term for the fermions, but in addition there might be a parity-even
mass, which breaks chirality\cite{HoPa93,KEIT95}. Both numerical and
analytical studies of the truncated Dyson--Schwinger (DS) equation for
the fermion propagator show that there is a first-order phase
transition\cite{KMpr95}. This is quite remarkable and in contrast to
the infinite-order phase transition one finds in pure
QED3\cite{qed3gen,lat}.

In this paper, based on Ref.\citen{KMpr95} in which more details are
given, we show how such a first-order transition arises, and we also
compare some different approximation schemes. In the next section, we
introduce the formalism we are using. In Sec.~\ref{secDMG} we discuss
the DS equation and the truncation schemes we are using, and how this
approach leads to dynamical mass generation in pure QED3. Next, we
give the results we find in quenched QED3 with CS term. In
Sec.~\ref{secIVP}, we include the effects of the vacuum polarization,
and finally we give some concluding remarks in Sec.~\ref{secCon}.

\section{Formalism}
\label{secF}

The Lagrangian in Euclidean space is
\begin{eqnarray}
{\cal L} &=& \bar\psi(i\!\not{\!\partial} + e\!\not{\!\! A}
             - m_e - \tau m_o )\psi
        + {\textstyle{1\over4}} F_{\mu\nu}^2
        + {\textstyle{1\over2}} i \theta \epsilon_{\mu\nu\rho}
                    A_\mu \partial_\nu A_\rho
        + {\cal L}_{\hbox{gauge fixing}} \,,
\label{lag}
\end{eqnarray}
with the dimensionful parameter $\theta$ determining the strength of
the CS term. We use four-component spinors for the fermions, and the
matrix $\tau$ is defined in such a way that the term $m_o \bar\psi\tau
\psi$ is odd under a parity transformation\cite{KEIT95,SW88CCW91}. Also the
CS term is odd under a parity transformation, the other terms in the
Lagrangian are parity invariant.

The inverse full fermion propagator can be written as
\begin{eqnarray}
 S^{-1}(p) &=& A_e(p) \not{\!p} + A_o(p)\tau \not{\!p}
                        - B_e(p) - B_o(p)\tau \,.
\end{eqnarray}
The functions $A(p)$ and $B(p)$ are scalar functions of $p^2$, and
their bare values are $A_e = 1$, $A_o = 0$, $B_e = m_e$, and $B_o =
m_o$. For convenience, we will introduce another decomposition, using
the projection operators $\chi_\pm \equiv \frac{1}{2}(1\pm\tau)$.
Defining $A_\pm \equiv A_e \pm A_o$ and $B_\pm \equiv B_e \pm B_o$,
this leads to
\begin{eqnarray}
 S(p)  &=&
 - \frac{A_+(p) \not{\!p} + B_+(p)}{A_+^2(p)p^2 + B_+^2(p)} \chi_+
 - \frac{A_-(p) \not{\!p} + B_-(p)}{A_-^2(p)p^2 + B_-^2(p)} \chi_-  \,.
\end{eqnarray}

With a four-dimensional representation for the $\gamma$-matrices, we
can define chirality similar as in four-dimensional QED. Without an
explicit mass $m_e$ for the fermions, the Lagrangian is chirally
symmetric, but the parity-even mass $m_e$ breaks this symmetry. Note
that the other mass, $m_o$, is chirally invariant. Just as in pure
QED, the chiral symmetry can be broken dynamically due to
nonperturbative effects: starting without a mass term for the fermions
in the Lagrangian, there might be a mass generated by the dynamics.
This can be studied using the fermion DS equation with the explicit
mass $m_e$ equal to zero. We also set the bare parity-odd mass, $m_o$,
equal to zero.

The gauge boson has not only a transverse and longitudinal part, but
also a parity-odd part, proportional to $\epsilon_{\mu\nu\rho}q_\rho$,
due to the explicit CS term in the Lagrangian. In a general covariant
gauge, the full gauge boson propagator can be written as
\begin{eqnarray}
 D_{\mu\nu}(q) &=&
 D^T(q^2)\left(\delta_{\mu\nu} - \frac{q_\mu q_\nu}{q^2}\right)
                   + D^O(q^2)\epsilon_{\mu\nu\rho}\frac{q_\rho}{|q|}
                   + a \frac{q_\mu q_\nu}{q^4}  \,,
 \label{Dprop}\\
 D^T(q^2) &=& \frac{q^2 - \Pi^T(q)}
                     {(q^2-\Pi^T(q))^2 + (\Pi^O(q) - \theta |q|)^2} \,,
 \label{Dtrans}\\
 D^O(q^2) &=& \frac{\Pi^O(q) - \theta |q|}
                     {(q^2-\Pi^T(q))^2 + (\Pi^O(q) - \theta |q|)^2} \,,
 \label{Dodd}
\end{eqnarray}
where $\Pi^T$ and $\Pi^O$ are the transverse and parity-odd part of
the vacuum polarization $\Pi_{\mu\nu}$. These parts can be projected
out, and this leads to
\begin{eqnarray}
 \Pi^T(q) &=& -2 N\, e^2  \int\!\frac{{\rm d^3}k}{(2\pi)^3}
   \Big( k^2 - 2 k\cdot q - 3 (k\cdot q)^2/q^2\Big)
   \times \nonumber \\
   && \left\{ \frac{A_+(k+q)A_+(k)}
   {\big((k+q)^2 A_+^2 + B_+^2\big)\big(k^2 A_+^2 + B_+^2\big)}
   + \Big[ A_+,B_+ \longrightarrow A_-, B_- \Big] \right\} \,,
 \label{Pitrans}\\
 \Pi^O(q) &=& \frac{4 N\, e^2}{|q|}\int\!\frac{{\rm d^3}k}{(2\pi)^3}
    \; q \cdot (k+q)  \times \nonumber \\
   && \left\{\frac{A_+(k+q) B_+(k)}
                 {((k+q)^2 A_+^2 + B_+^2)(k^2 A_+^2 + B_+^2)}
               - \Big[ A_+,B_+ \longrightarrow A_-, B_- \Big] \right\}\,,
 \label{Piodd}
\end{eqnarray}
using a bare vertex. Note that the longitudinal part of the vacuum
polarization is zero because of gauge invariance.

\section{Dynamical Mass Generation}
\label{secDMG}

In order to study dynamical mass generation for the fermions, we use a
truncation of the DS equation for the fermion propagator.

\subsection{Dyson--Schwinger Equation}

The set of DS equations forms a nonperturbative, exact set of
equations for Green's functions, and for the fermion propagator it
reads
\begin{eqnarray}
  S^{-1}(p) =    S^{-1}_0(p) -
     e^2\!\int\!\!\frac{{\rm d}^3k}{(2\pi)^3}
     \gamma_\mu S(k) \Gamma_\nu(p,k) D_{\mu\nu}(p-k) \,.
\end{eqnarray}
The problem in solving this equation is the the unknown full vertex
function $\Gamma_\nu(p,k)$, and the vacuum polarization
$\Pi_{\mu\nu}(q)$ which appears in the gauge boson propagator. In
principle, we could write down DS equations for these Green's
functions as well, but the DS equation for the vertex involves a
four-point Green's function, and so on. For practical purposes, we
have to truncate the set of DS equations.

In this paper, we consider some different truncation schemes. First,
we take the most simple truncation, namely replacing the full vertex
and the full gauge boson by the bare ones, see Sec.~\ref{secQQED}.
This is usually referred to as the quenched ladder approximation. We
also consider a truncation based on the $1/N$ expansion.

\subsection{1/N Expansion}

A popular approximation scheme in QED3 with $N$ fermion flavors is the
$1/N$ expansion\cite{expansion}. The coupling constant $e^2$ has the
dimension of mass, and we can use the large $N$ limit in such a way
that $e^2\downarrow 0$ and the product $N\,e^2$ remains fixed:
$Ne^2=8\alpha$ with $\alpha$ fixed. In QED3, the most commonly used
truncation scheme of the DS equation is based on this $1/N$ expansion.

In this truncation scheme, the one-loop vacuum polarization has to be
taken into account, since it is of order one: there are $N$ fermion
loops, and each loop is proportional to $1/N$. The full vertex is
replaced by the bare one, because that is the leading order
contribution in $1/N$; all vertex corrections are of down by a factor
of $1/N$. This truncation scheme gives two sets of two coupled
integral equations for $A_+$ and $B_+$, respectively for $A_-$ and
$B_-$
\begin{eqnarray}
A_\pm(p) &=& 1 + \frac{e^2}{p^2}
 \int\!\frac{{\rm d^3}k}{(2\pi)^{\rm 3}}
            \frac{2}{k^2 A_\pm^2(k) + B_\pm^2(k)} \times
\nonumber \\ &&
           \Bigg( A_\pm(k) \bigg(
  \Big(D^T(q) - \frac{a}{q^2}\Big) \frac{q\cdot k \; q\cdot p}{q^2}
              + \frac{a \, k\cdot p}{2 q^2}\bigg)
          \pm B_\pm(k) D^O(q) \frac{p\cdot q}{|q|} \Bigg) \,,
\label{Aeq}\\
B_\pm(p) &=& e^2 \int\!\frac{{\rm d^3}k}{(2\pi)^{\rm 3}}
               \frac{1}{k^2 A_\pm^2(k) + B_\pm^2(k)} \times
\nonumber\\ &&
           \Bigg(B_\pm(k)\Big(2 D^T(q) + \frac{a}{q^2} \Big)
  \mp 2 A_\pm(k) D^O(q) \frac{k\cdot q}{|q|}\Bigg)  \,.
\label{Beq}
\end{eqnarray}
where $D^T$ and $D^O$ are given by Eqs.~(\ref{Dtrans}) and
(\ref{Dodd}); $a$ is the gauge parameter. Note that the equations for
$A_+$ and $B_+$ are coupled to the ones for $A_-$ and $B_-$ through
the vacuum polarization only, at least in this truncation scheme. The
quenched ladder approximation can be recovered by taking a bare gauge
boson propagator.

Of course, we have to make sure that we satisfy at least approximately
the Ward--Takahashi identity. One of the consequences of the
Ward--Takahashi identity is that the vertex renormalization and the
wave-function renormalization have to be equal. This means that if we
are using a bare vertex approximation, we have to make sure that $A_+
= A_- = 1$, at least approximately. To achieve this, we can use the
argument that (formally) the wave-function renormalization is $1 +
{\cal O}(1/N)$, see Eq.~(\ref{Aeq}) with $e^2=8\alpha/N$, so to
leading order in $1/N$ we can neglect the wave-function
renormalization. Alternatively, we could use a suitable nonlocal
(momentum dependent) gauge function $a(q)$, which makes the
wave-function renormalization equal to one\cite{Si90KuMi92}.

\subsection{Pure QED3}

In QED3 without CS term, dynamical breaking of chiral symmetry has
been studied extensively\cite{qed3gen,lat}. It is known that parity is
not broken dynamically\cite{noparitybroken}, so the parity-odd
functions $A_o$ and $B_o$ are zero. Using either a nonlocal gauge such
that $A(p) = 1$, or just neglecting the wave-function renormalization,
the set of Eqs.~(\ref{Aeq}) and (\ref{Beq}) reduces to one equation
for the dynamical mass function $m(p) = B(p)/A(p)$
\begin{eqnarray}
 m(p) &=& \frac{8\,\alpha}{N} \int\!\frac{{\rm d^3}k}{(2\pi)^{\rm 3}}
 \frac{m(k)}{k^2 + m^2(k)} \left(2 D^T(q) + \frac{a}{q^2} \right) \,.
\label{meq}
\end{eqnarray}

With the one-loop vacuum polarization of bare massless fermions,
$\Pi^T=-\alpha q$, there exists a critical number of fermion flavors,
$N_c$, above which there is no chiral symmetry breaking; for $N \leq
N_c$, there is a dynamically generated fermion mass\cite{qed3gen,lat}.
Close to the critical number, the value of the mass function at the
origin behaves like
\begin{eqnarray}
  m(0) &\sim& \alpha \exp{\left(\frac{-2\pi}{\sqrt{N_c/N -1}}\right)} \,,
\end{eqnarray}
corresponding to an infinite-order phase transition. This behavior is
obtained both numerically and analytically, using bifurcation theory.
Recently, this critical behavior has been confirmed by numerical
calculations including the effects of the generated fermion mass on
the vacuum polarization\cite{GuHaRe95}. The critical number $N_c$
depends on the truncation and gauge: in the Landau gauge,
$N_c={32}/{\pi^2}$, whereas in the nonlocal gauge, one finds
$N_c={128}/{(3\pi^2)}$. Apart from this difference of a factor of
$\frac{4}{3}$, the results in Landau and nonlocal gauge agree with
other.

\section{Results with a Chern--Simons Term in Quenched QED}
\label{secQQED}

One important difference with pure QED, is that with a CS term, there
is no trivial solution $B_\pm = 0$; there is always the
chirally-symmetric, parity-odd solution $B_+(p)= -B_-(p)= B_o(p)$,
with $B_e(p) = 0$. This solution can also be found using perturbation
theory. Dynamical chiral symmetry breaking can only occur if there is
a solution with $B_+ \neq -B_-$, in such a way that $B_e(p) =
\frac{1}{2}(B_+(p) + B_-(p)) $ is nonzero. Furthermore, in the
ultraviolet region, the CS term dominates, leading to a behavior of
the mass function at large momenta like $B_\pm(p)\sim \pm\theta/p$,
whereas the dynamical mass function in pure QED3 falls off like
$B(p)\sim 1/p^2$.

\subsection{Numerical Results}

In quenched QED3 with a CS term, the equations for $A_+$ and $B_+$
decouple from the equations for $A_-$ and $B_-$, as can be seen from
Eqs.~(\ref{Aeq}) and (\ref{Beq}) with a bare gauge boson. By choosing
the Landau gauge, $a = 0$, we find that $A = 1$ exactly in pure QED,
and for small values of $\theta$ (small compared to the mass scale
$e^2$), $A_\pm \simeq 1$, so we neglect the wave-function
renormalization, and consider the equations for $B_+$ and $B_-$ only.

\begin{figure}
 \epsfysize = 6.4cm
 \centerline{\epsffile{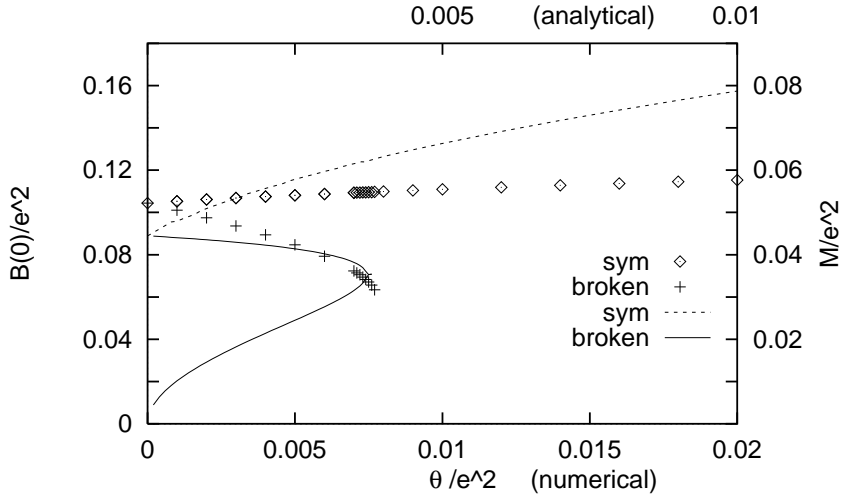}}
 \caption{$B(0)$ versus $\theta$, both numerically (diamonds: symmetric
          solution, and plusses: broken solution) and analytically
          (dashed: symmetric, and solid line: broken solution).}
 \label{figBvstquen}
\end{figure}

Numerically, we find the following solutions:
\begin{itemize}
\item
$B_+(p) = - B_-(p)$,
the symmetric solution with $B_e(p) = 0$;
\item
$B_+(0) \simeq \tilde B_-(0) > 0$,
with $B_e(p) \neq 0$ and $B_o(p) = {\cal O}(\theta)$.
\end{itemize}
The last solution exists for small values of $\theta$ only, and
corresponds to dynamical breaking of chiral symmetry. For large values
of $\theta$, we only find the chirally-symmetric solution. Between
these two regions there is a critical value of $\theta$ for dynamical
chiral symmetry breaking. As can be seen from Fig.~\ref{figBvstquen},
the second solution, $\tilde B$, decreases for increasing $\theta$,
and beyond some critical value of $\theta$ we cannot find this
solution any longer, but at this critical value, $\tilde B(0)$ nor the
chiral condensate go to zero. This clearly indicates that it is a
discontinuous, first-order phase transition.
%
%

\subsection{Analytical Results}

Although the numerical results indicate that it is a first-order phase
transition, it is very difficult to establish such a discontinuous
transition numerically. In order to solve the equations analytically,
we make an expansion in $\theta$, keeping the linear terms only, and
perform the angular integrals analytically. Next, we use an Ansatz for
the solution based on the leading ultraviolet behavior of the mass
functions $ B_\pm(p) \simeq \pm {e^2\,\theta}/{(\pi^2\,p)}$, but with
a finite value at the origin. Qualitatively, this Ansatz behaves quite
similar to the numerical solution we have found in the previous
section. This leads to a boundary condition for $M \equiv B_+(0)$
\begin{eqnarray}
 M^3 - \frac{e^2}{\pi^2} \bigg(
 \frac{M^3}{|M|}
 \Big({\textstyle\frac{\pi}{4}}- {\textstyle\frac{1}{2}}\ln 2\Big)
 + \frac{e^2\,\theta}{2\pi^2}(1+\ln 2) +
  \frac{\pi\,\theta\,M}{2} \bigg) &=& 0 \,.
\end{eqnarray}
\begin{wrapfigure}{r}{6.8cm}
 \epsfysize = 4.9cm
 \centerline{\epsffile{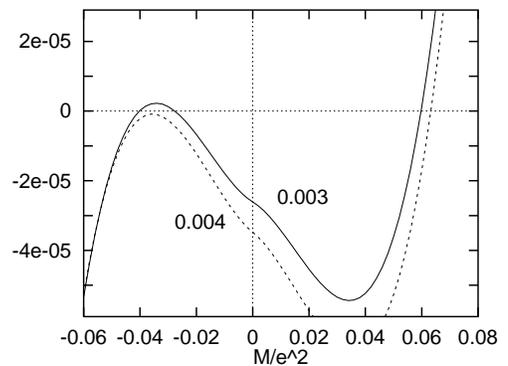}}
 \caption{Boundary condition for $M$ at $\theta = 0.003$
          and $\theta = 0.004$, just below and above $\theta_c$.}
 \label{figbcquen}
\end{wrapfigure}
%
%
This cubic equation in $M$ has either one or three solutions for given
$\theta$, as can be seen from Fig.~\ref{figbcquen}: for small
(nonzero) values of $\theta$ there are three solutions, and there is
dynamical breaking of chiral symmetry, whereas for large values of
$\theta$ there is only the chirally symmetric solution. From this
figure it is clear that it is indeed a discontinuous transition, at
the critical coupling $\theta_c$ the ``second'' solution $\tilde B$
does not go to zero, nor does the chiral condensate. So both our
numerical and analytical calculations are in good qualitative
agreement, and show that there is a first-order chiral phase
transition. Quantitatively there is about a factor of two difference
between our numerical and analytical results
\begin{eqnarray*}
\begin{array}{rclrcl}
    \theta_c/e^2 &=& 0.00371 {\hbox{   (an.)}}
  & \theta_c/e^2 &\simeq&   0.008 {\hbox{   (num.)}}\\
  \tilde B_c/e^2 &=& 0.0352
  &\tilde B_c/e^2 &\simeq&   0.06 \;\;,
\end{array}
\end{eqnarray*}
caused by the approximations we made in solving the equation for
$B_\pm$ analytically.

\section{Influence of the Vacuum Polarization}
\label{secIVP}

Next, we consider the case of $N$ fermion flavors, and include the
effects of the vacuum polarization. Since we know already that there
is a critical number of fermion flavors for dynamical chiral symmetry
breaking in pure QED, we expect the same to be true in the presence of
a (small) CS term. We first use the one-loop vacuum polarization with
bare massless fermions.

\subsection{One-loop Vacuum Polarization}

With massless fermions, there is no parity-odd part of the vacuum
polarization, as can be seen from Eq.~(\ref{Piodd}), and the
transverse part is simply $ \Pi_T(q)=-\alpha q$, with $\alpha=Ne^2/8$.
Thus the transverse and parity-odd part of the gauge boson become
\begin{eqnarray}
 D^T(q) = \frac{q^2 + \alpha\,q}
    {q^2\left( (q + \alpha)^2 + \theta^2\right)}  \,,
\qquad&&\qquad
 D^O(q) = \frac{ - \theta q}
     {q^2\left( (q + \alpha)^2 + \theta^2\right)}  \,.
\end{eqnarray}
Again, the equation for $A_+$ and $B_+$ decouple from the ones for
$A_-$ and $B_-$, and it should be noted that if we have a solution
$(A_+,B_+)$ of the ``plus'' equations, then the combination
$(A_+,-B_+)$ is automatically a solution of the ``minus'' equations.

We used both the Landau gauge and a nonlocal gauge\cite{KEIT95} to
keep $A_\pm \simeq 1$. In Landau gauge, $A_\pm$ will in general not be
equal to one, but we nevertheless set it equal to one, in order to be
consistent with the bare vertex approximation. Most of the results we
present here, are obtained in the nonlocal gauge; in Landau gauge the
results are quite similar, the main difference being a scaling factor
of $\frac{4}{3}$ in $N$, the number of flavors.

\begin{figure}[t]
 \epsfysize = 5.2cm
 \centerline{\epsffile{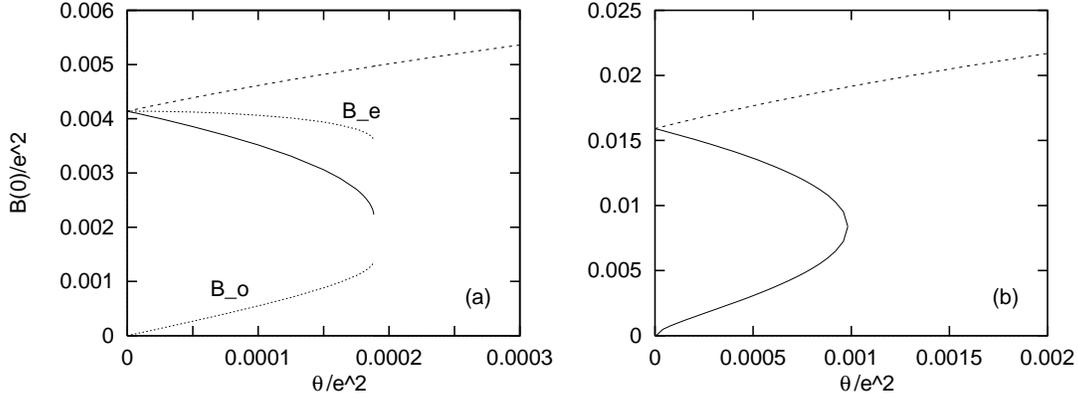}}
 \caption{$B(0)$ (dashed line) and $\tilde B(0)$ (solid line) vs.
          $\theta$ at $N=2$ in nonlocal gauge:
          (a) numerically and (b) analytically.
          For completeness, we also included $B_e(0)$ and $B_o(0)$ in (a).}
 \label{fig1loop}
\end{figure}

Numerically, we have found results similar to the quenched case, at
least below $N_c(\theta=0)$, the critical number of fermion flavors in
pure QED. For fixed $\theta$, we find a critical $N_c(\theta)$, above
which there is no chiral symmetry breaking, and also for small $N$,
there is a critical $\theta_c(N)$ for chiral symmetry breaking. Our
numerical results indicate a discontinuous (first-order) phase
transition, as can be seen from Fig.~\ref{fig1loop}(a).
%
%

We also solved the equation analytically, in order to show that the
phase transition is indeed first order.  Some standard approximations
lead to the general solution
\begin{eqnarray}
  B_\pm(p) & = & M_\pm \, _2F_1(a_+,a_-,\frac{3}{2};-p^2/M_\pm^2) \,,
\end{eqnarray}
with a boundary condition for $M_\pm \equiv B_\pm(0)$
\begin{eqnarray}
\pm \frac{8\theta}{\pi^2 N}\left( 1 + \frac{16}{9\pi^2 N }\right) \;=\;
  M_\pm \; {}_2F_1(a_+,a_-,{\textstyle\frac{1}{2}};-\alpha^2/M_\pm^2)
 \nonumber \\
 - \frac{8}{3\pi^2 N} M_\pm \;
 {}_2F_1(a_+,a_-,{\textstyle\frac{3}{2}};-\alpha^2/M_\pm^2) \,,
\end{eqnarray}
with $a_\pm = \frac{1}{4} \pm \frac{1}{4}i\sqrt{(N_c/N - 1)}$; the
only difference between the Landau and nonlocal gauge is the critical
number: $N_c={32}/{\pi^2}$ in Landau and ${128}/{(3\pi^2)}$ in
nonlocal gauge.

This boundary condition gives us $M$ as function of $N$ and $\theta$,
and is plotted
\begin{wrapfigure}{r}{6.8cm}
 \epsfysize = 5.2cm
 \centerline{\epsffile{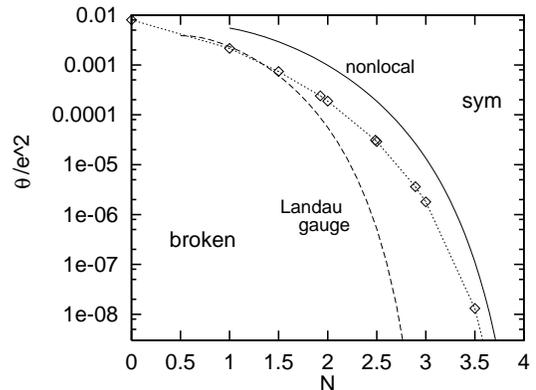}}
 \caption{Phase diagram; the diamonds are
          numerical results in nonlocal gauge.}
 \label{figNtdiagram}
\end{wrapfigure}
%
%
in Fig.~\ref{fig1loop}(b). As can be seen from this figure, there is a
critical $\theta_c$, beyond which the chirally symmetric solution is
the only one. The generated mass $\tilde B$ does not go to zero at the
critical point, corresponding to a first-order phase transition. Also
in the other direction, increasing $N$ at fixed $\theta$, we find a
discontinuous transition, both numerically and
analytically\cite{KMpr95}. So both our numerical and analytical
results are in qualitative agreement with each other; also both Landau
and nonlocal gauge are in good qualitative agreement.

In Fig.~\ref{figNtdiagram} we have shown the critical line in the
$(N,\theta)$ plane.

\subsection{Full Vacuum Polarization}

The above results are obtained by using the one-loop vacuum
polarization with massless fermions. Now one can of course raise the
question what is the influence of the generated mass on the vacuum
polarization. In pure QED3, the generated mass has no effect on the
critical number of fermion flavors, nor on the behavior of $m(0)$ near
the phase transition\cite{GuHaRe95}, but it does have a crucial effect
on the confining properties of this model\cite{BuPrRo92} and on
behavior of the mass function in the complex momentum
plane\cite{Ma95}.

\begin{figure}
 \epsfysize = 5.2cm
 \centerline{\epsffile{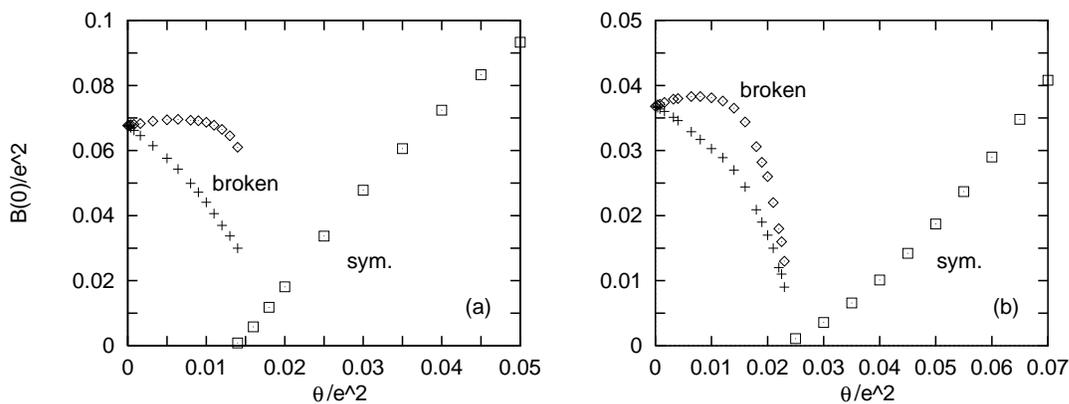}}
 \caption{$\tilde B_+(0)$ (diamonds), $\tilde B_-(0)$ (plusses),
          and $B_+(0)$ (squares) vs. $\theta$ at
          (a) $N=0.5$ and (b) $N=1$.}
 \label{figfulvacpol}
\end{figure}

Using the vacuum polarization with the dynamical fermion mass, we have
both a transverse and a parity-odd part of the vacuum polarization,
and the equations for the ``plus'' and ``minus'' functions become
coupled. For simplicity, we use the Landau gauge, and put $A_\pm = 1$
by hand, so we have to deal with a set of four coupled nonlinear
integral equations for $B_+,B_-,\Pi^T, \Pi^O$. Our numerical results
show that there is dynamical chiral symmetry breaking at small values
of $\theta$ and $N$, just as in the previous case. For large values of
$\theta$ and $N$, we only find the chirally symmetric solution. In
Fig.~\ref{figfulvacpol} we have shown our first results. Sofar, the
numerical results indicate that the phase transition is again
discontinuous, but we need more accurate results close to the critical
parameters before we can establish a first-order phase transition; it
might be a very sharp second order transition. More detailed
calculations of the behavior near the critical line should be done,
both numerically and analytically.
%
%

Finally, we should note that an important feature of vacuum
polarization is that the parity-odd part is {\em linear} in the
generated fermion mass $B$. This linearity turns out to be important:
because of this, the symmetric solution, which is parity-odd, vanishes
for vanishing $\theta$. This is in agreement with other analyses,
showing that there is no dynamical breaking of parity in pure
QED3\cite{noparitybroken}, but in contrast to the results obtained
with the one-loop vacuum polarization of massless fermions, see
Fig.~\ref{fig1loop}.

\section{Conclusion}
\label{secCon}

In QED3 with $N$ fermion flavors, we have solved the DS equation in
the presence of an explicit CS term. Without this CS term, there is
dynamical chiral symmetry breaking if the number of fermion flavors is
below some critical number $N_c$. The chiral phase transition at
$N=N_c$ is of infinite order; the order parameter and the generated
mass vanish continuously at the phase transition.

In the presence of an explicit CS term for the gauge field, there can
be both a chirally symmetric mass, and a mass which breaks the chiral
symmetry. Below the critical number $N_c$, we find that there is
chiral symmetry breaking for small values of $\theta$, just as in pure
QED3. There is a critical line in the $(N,\theta)$ plane, beyond which
the chirally symmetric solution is the only solution. At this critical
line, the order parameter and the generated mass do {\em not} vanish,
corresponding to a first-order phase transition. Both numerically and
analytically we have found this discontinuous transition both in
quenched QED3, and using the one-loop vacuum polarization. First
results including the generated mass in the vacuum polarization
indicate the same behavior.

Note that there is no dynamical breaking of parity. In our analysis,
this can be seen from the fact that the chirally-symmetric, parity-odd
solution vanishes for vanishing $\theta$. One has to take into account
the effect of the generated fermion mass on the vacuum polarization in
order to get this result.

\section*{Acknowledgements}

Most of this work has been done in collaboration with K.-I.~Kondo.
I would also like to thank Y.~Kim, K.~Yamawaki and D.K.~Hong for
interesting discussions. This work is financially supported by
the Japanese Society for the Promotion of Science.


\begin{thebibliography}{99}

\bibitem{qed3gen}
 R.D. Pisarski, Phys. Rev. {\bf D29} (1984), 2423;
 T.W. Appelquist, M.J. Bowick, D. Karabali, and L.C.R. Wijewardhana,
 Phys. Rev. {\bf D33} (1986), 3704;
 T.W. Appelquist, D. Nash, and L.C.R. Wijewardhana,
 Phys. Rev. Lett. {\bf 60} (1988), 2575;
 D. Nash, Phys. Rev. Lett. {\bf62} (1989), 3024.

\bibitem{lat}
 E. Dagotto, A. Koci\'c, and J.B. Kogut,
 Phys. Rev. Lett. {\bf 62} (1989), 1083;
 Nucl. Phys. {\bf B334} (1990), 279.

\bibitem{cs}
 S. Deser, R. Jackiw, and S. Templeton,
 Ann. Phys. {\bf 140} (1982), 372.

\bibitem{KMpr95}
 K.-I. Kondo and P. Maris, Phys. Rev. Lett. {\bf 74} (1995), 18;
 Phys. Rev. {\bf D52} (1995), 1212.

\bibitem{Ko95}
 K.-I. Kondo, {\it First and Second Order Phase Transition
 in Maxwell--Chern--Simons Theory Coupled  to Fermions},
 CHIBA-EP-89, hep-ph/9509345.

\bibitem{HoPa93}
 D.K. Hong and S.H. Park, Phys. Rev. {\bf D47} (1993), 3651.

\bibitem{KEIT95}
 K.-I. Kondo, T. Ebihara, T. Iizuka and E. Tanaka,
 Nucl. Phys. {\bf B434} (1995), 85.

\bibitem{SW88CCW91}
 G.W. Semenoff and L.C.R. Wijewardhana,
 Phys. Rev. Lett. {\bf 62} (1988), 2633;
 M. Carena, T.E. Clark, and C.E.M. Wagner,
 Phys. Lett. {\bf B259} (1991), 128; Nucl. Phys. {\bf B356} (1991), 117.

\bibitem{expansion}
 R. Jackiw and S. Templeton, Phys. Rev. {\bf D23} (1981), 2291;
 T.W. Appelquist and R.D. Pisarski, Phys. Rev. {\bf D23} (1981), 2305;
 T.W. Appelquist and U. Heinz, Phys. Rev. {\bf D23} (1981), 2169.

\bibitem{Si90KuMi92}
 E.H. Simmons, Phys. Rev. {\bf D42} (1990), 2933;
 T. Kugo and M.G. Mitchard, Phys. Lett. {\bf B282} (1992), 162.

\bibitem{noparitybroken}
 T.W. Appelquist, M.J. Bowick, D. Karabali, and L.C.R. Wijewardhana,
 Phys. Rev. {\bf D33} (1986), 3774;
 C. Vafa and E. Witten, Commun. Math. Phys. {\bf 95} (1984), 257.

\bibitem{GuHaRe95}
 V.P. Gusynin, A.H. Hams, and M. Reenders,
 {\it (2+1)-Dimensional QED with Dynamically Massive Fermions
 in the Vacuum Polarization}, UG-10-95, hep-ph/9509380.

\bibitem{BuPrRo92}
 C.J. Burden, J. Praschifska, and C.D. Roberts,
 Phys. Rev. {\bf D46} (1992), 2695.

\bibitem{Ma95}
 P. Maris, {\it Confinement and Complex Singularities in QED3},
 DPNU-95-20, hep-ph/9508323.

\end{thebibliography}
\end{document}